\documentclass[pra,twocolumn,preprintnumbers,amsmath,amssymb,superscriptaddress]{revtex4-1}

\usepackage[ansinew]{inputenc}
\usepackage[T1]{fontenc}
\usepackage{ae,aecompl}
\usepackage[english]{babel}
\usepackage[dvips]{graphicx}
\usepackage{dsfont}
\usepackage{amsmath}
\usepackage{pifont}
\usepackage{enumerate,amsthm,amsmath,amssymb,color,float}
\usepackage{enumerate,color,float}

\usepackage[caption=false]{subfig}

\newcommand{\op}[1]{\mbox{$\hat{#1}$}}
\newcommand{\ket}[1]{\vert#1\rangle}
\newcommand{\bra}[1]{\langle#1\vert}

\begin{document}

\title{Noiseless Linear Amplification and Quantum Channels}
\author{R\'emi Blandino} 
\email{r.blandino@uq.edu.au}
\affiliation{Centre for Quantum Computation and Communication Technology, School of Mathematics and Physics, University of Queensland, St Lucia Queensland 4072, Australia}
\affiliation{Laboratoire Charles Fabry, Institut d'Optique, 
CNRS, Universit\'e Paris-Sud, Campus Polytechnique, RD 128, 91127 Palaiseau cedex, France}
\author{Marco Barbieri}
\affiliation{Dipartimento di Scienze, Universit\`a degli Studi Roma 3, Via della Vasca Navale 84, 00146, Rome, Italy}
\affiliation{Clarendon Laboratory, Department of Physics, University of Oxford, Parks Road, OX1 3PU, Oxford, UK}
\author{Philippe Grangier}
\affiliation{Laboratoire Charles Fabry, Institut d'Optique, 
CNRS, Universit\'e Paris-Sud, Campus Polytechnique, RD 128, 91127 Palaiseau cedex, France}
\author{Rosa Tualle-Brouri} 
\affiliation{Laboratoire Charles Fabry, Institut d'Optique, 
CNRS, Universit\'e Paris-Sud, Campus Polytechnique, RD 128, 91127 Palaiseau cedex, France}
\affiliation{Institut Universitaire de France, 103 boulevard St. Michel, 75005, Paris, France}

\begin{abstract}
The employ of a noiseless linear amplifier (NLA) has been proven as a useful tool for mitigating imperfections in quantum channels. Its analysis is usually conducted within specific frameworks, for which the set of input states for a given protocol is fixed. Here we obtain a more general description by showing that a noisy and lossy Gaussian channel followed by a NLA has a general description in terms of effective channels. This has the advantage of offering a simpler mathematical description, best suitable for mixed states, both Gaussian and non-Gaussian. We investigate the main properties of this effective system, and illustrate its potential by applying it to loss compensation and reduction of phase uncertainty. 
\end{abstract}
\maketitle


\section{Introduction}

Deterministic phase-insensitive quantum amplifiers, which amplify equally any quadrature of light, are fundamentally limited by quantum physics and must add a minimal amount of quantum noise \cite{caves_quantum_2012}. A Noiseless Linear Amplifier (NLA), on the other hand, can in theory achieve a phase-insensitive amplification which does not add any noise, and more surprisingly which does not amplifies the quantum noise, but at the expense of a probabilistic transformation \cite{ralph_nondeterministic_2008}. 

Noiseless linear amplification have been actively studied from various perspectives. The first one concerns the implementation of the NLA itself, since a perfect noiseless amplification can only occur with a zero probability of success \cite{pandey_quantum_2013}. However, one can obtain an output with a very high fidelity and non-zero probability if the output state if well approximated within a $N$ dimensional Hilbert space. Increasing this value, and hence the working range of the approximate NLA, inevitably decreases the probability of success. Several methods have been proposed and experimentally realized to implement an approximate NLA \cite{ralph_nondeterministic_2008,fiurasek_engineering_2009,menzies_noiseless_2009,ferreyrol_implementation_2010,ferreyrol_experimental_2011,xiang_heralded_2010,zavatta_high-fidelity_2011,usuga_noise-powered_2010}. Some implementations have also been proposed in order to increase the probability of success \cite{dunjko_truly_2012}, or to avoid the use of non-Gaussian resources \cite{marek_coherent-state_2010,partanen_noiseless_2012} when restricted to the amplification of coherent states. The NLA has also been studied from a more abstract point of view \cite{walk_nondeterministic_2013}, and with a focus on optimal design and probability of success  \cite{mcmahon_optimal_2014,chiribella_optimal_2013}.

The second perspective has focused on the use of the NLA for various applications, such as quantum information protocols or quantum state preparation \cite{gagatsos_heralded_2014}, either considering a perfect NLA as a theoretical limit, or an approximated one. The NLA has for instance been shown to be useful in quantum key distribution, for continuous-variable \cite{blandino_improving_2012,walk_security_2013,fiurasek_gaussian_2012} as well as discrete-variable  \cite{kocsis_heralded_2013,gisin_proposal_2010,osorio_heralded_2012}. It can also be used for loss suppression \cite{micuda_noiseless_2012,meng_noiseless_2012}, Bell-inequality violation \cite{brask_bell_2012,torlai_violation_2013}, entanglement distillation \cite{ralph_nondeterministic_2008,yang_continuous-variable_2012}, quantum cloning \cite{muller_probabilistic_2012}, phase-insensitive squeezing \cite{gagatsos_probabilistic_2012}, or error correction  \cite{ralph_quantum_2011}.

A promising result towards a practical use of the NLA is the possibility to implement it virtually, using only post-selection \cite{walk_security_2013,fiurasek_gaussian_2012}, as experimentally demonstrated for entanglement distillation
\cite{chrzanowski_measurement-based_2014}.

\begin{figure}[t]
\subfloat[]{\includegraphics[width= \columnwidth]{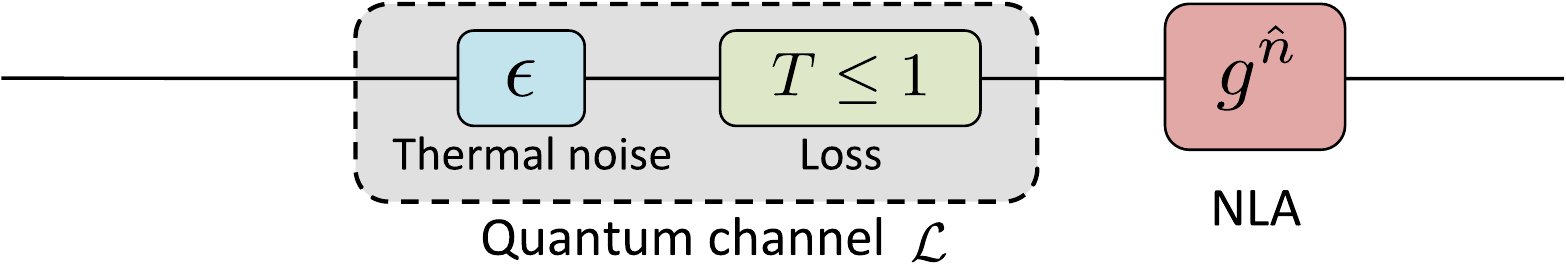}} \\
\subfloat[]{\includegraphics[width= \columnwidth]{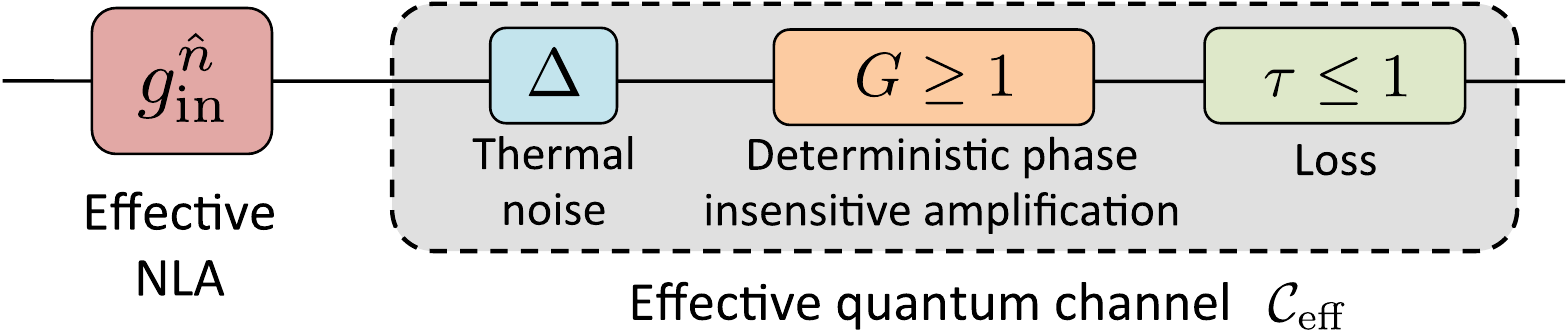}}
\caption{(Color online) Two equivalent systems, up to a global state-independent normalization factor: (a) NLA placed after a linear symmetric lossy and noisy Gaussian quantum channel. (b) Effective NLA placed before an effective linear symmetric Gaussian channel, composed of input noise, deterministic phase insensitive amplification and loss. See main text for more details.}
\label{ampli_deterministe_phaseIndependant}
\end{figure}

Most of the analyses mentioned above start by considering specific protocols, hence address a specific class of input states, such as coherent states. This is an effective approach, but clearly lacks generality, in particular when one is interested in using the NLA with non-Gaussian states. In this paper we present a generalization which allows to describe the NLA acting after a Gaussian channel as an effective channel. The usefulness of such a description is twofold. First, the noiseless amplification from the effective system is usually simpler to compute, especially if the input state is pure, as assumed in most protocols. Second, it gives a physical insight in the transformation produced by the noiseless amplification, and allows to find new protocols and applications.

The outline is as follows: in the first part, we show that a linear symmetric lossy and noisy Gaussian quantum channel followed by a NLA produces the same transformation as an effective NLA of different gain, followed by an effective linear symmetric Gaussian quantum channel, as shown in Fig. \ref{ampli_deterministe_phaseIndependant}. The effective quantum channel is then studied is detail. We analyze some behaviors and physical constraints on the effective parameters, and show that the effective channel can be reduced to a simpler one in some cases. 

In the second part, we use those results to present two new potential applications of the NLA. We first generalize the results of \cite{micuda_noiseless_2012}, and show that an exact loss reduction is achievable using the effective system. The second application concerns the `phase concentration' and the signal-to-noise ratio (SNR). We show that the addition of thermal noise can improve both of them, thanks to a non trivial behavior of the effective parameters.


\section{Effective system}

Let us start by introducing the main ideas leading to the effective system, while leaving the detailed calculation in the appendix. We consider a perfect NLA, described by an operator $\op{T}{=}g^{\hat{n}}$, which transforms a coherent state $\ket{\alpha}$ to
\begin{equation}
\op{T}\ket{\alpha}=e^{\frac{\vert \alpha \vert^2}{2}(g^2{-}1)}\ket{g \alpha}.
\label{amplification_coherent_state}
\end{equation}

We stress that our approach focuses on a linear state-independent regime, which is to be understood as a theoretical limit for any physical implementation of the NLA. In practice, a given physical implementation will act as a NLA only for a limited range of input states, which can however be made arbitrarily large, e.g. by increasing the number of stages in the quantum scissors scheme \cite{ralph_nondeterministic_2008}. 

The effective system is obtained by computing the output state with two methods, for an arbitrary input state. The first method corresponds to the quantum channel followed by the NLA.  The second methods corresponds to the effective system, where the input state is first noiselessly amplified, and then sent through an effective quantum channel. By comparing the two outputs, we can get the expressions of the parameters of the effective system such that the two transformations are equal.

\subsection{Effective parameters}

Let  $\op{\rho}_{\rm in}$ be an arbitrary quantum state, which we express using the $P$ function \cite{gerry_introductory_2005}:
\begin{align}
\op{\rho}_{\rm in}=\int \mathrm{d}^2\gamma \text{ } P_{\rm in}(\gamma)\ket{\gamma}\bra{\gamma} \label{decomposition_rho_in_NLA}
\end{align}
Note that $P_{\rm in}$ may in general be ill-behaved for non classical states, including squeezed and non-Gaussian states, however, we will not need its explicit expression, but simply use the linearity of the transformations in the coherent states basis to obtain the expression of the effective system. Several analytical and numerical tests have been performed to ensure the validity of this approach, by comparing the output states obtained with the direct system and with the effective system, for various Gaussian and non-Gaussian states. 

\subsubsection{Output state after the initial channel and the NLA}

The action of the initial channel $\mathcal{L}$ on the input state $\op{\rho}_{\rm in}$ can be described by a linear quantum operation $\mathfrak{L}$, which transforms a coherent state of mean amplitude $\gamma$ to a thermal state of parameter $\lambda_{\rm ch}$ and mean amplitude $\sqrt{T} \gamma$. As shown in the appendix \ref{appen1}, the action of a NLA on such a displaced thermal state produces another displaced thermal state
\begin{align}
\op{\sigma}(\gamma)=\op{D}\big(\tilde{g}\sqrt{T}\gamma\big)\op{\rho}_{\rm th}(g\lambda_{\rm ch})\op{D}^\dagger\big(\tilde{g}\sqrt{T}\gamma\big)
\label{displaced_thermal} 
\end{align}
of parameter $g \lambda_{\rm ch}$ and mean amplitude $\tilde{g} \sqrt{T} \gamma$, where the gain  $\tilde{g}$ is given by
\begin{align}
\tilde{g}=g\frac{1{-}\lambda_{\rm ch}^2}{1{-}g^2\lambda_{\rm ch}^2}. \label{tilde_g}
\end{align}

This allows us to obtain the output state $\op{\rho}_{\rm out}^{\rm NLA} $ produced by the system depicted in Fig. \ref{ampli_deterministe_phaseIndependant} (a),
\begin{align}
&\op{\rho}_{\rm out}^{\rm NLA} \propto \int \mathrm{d}^2\gamma \text{ } P_{\rm in}(\gamma)  \op{\sigma}(\gamma) e^{\vert \gamma \vert^2 T \frac{(g^2{-}1)(1{-}\lambda_{\rm ch}^2)}{1{-}g^2\lambda_{\rm ch}^2}}.
\label{etat_amplifie_apres}
\end{align}

\subsubsection{Output state after the effective system}

We now consider the case depicted in Fig. \ref{ampli_deterministe_phaseIndependant} (b), where a NLA of gain $g_{\rm in}$ is directly applied to the input state $\op{\rho}_{\rm in}$. Using again the decomposition \eqref{decomposition_rho_in_NLA}, the action of this NLA on a coherent state $\ket{\gamma}\bra{\gamma}$ can be directly obtained from \eqref{amplification_coherent_state}. In order to obtain the same exponential factor as in \eqref{etat_amplifie_apres}, $g_{\rm in}$ needs to satisfy
 \begin{align}
g^2_{\rm in}{-}1=T \frac{(g^2{-}1)(1{-}\lambda_{\rm ch}^2)}{1{-}g^2\lambda_{\rm ch}^2}.
\label{condition_gin}
\end{align}

We seek for a more general channel $\mathcal{C}_{\rm eff}$ after the effective NLA, as depicted in Fig. \ref{ampli_deterministe_phaseIndependant} (b). In the most general case, a deterministic linear symmetric Gaussian channel is composed of three elements: an addition of thermal noise $\Delta$ at its input; a deterministic phase-insensitive amplifier of intensity gain $G{\geq}1$, limited to the quantum noise; and a noiseless lossy channel of transmission $\tau{\leq}1$. As discussed below, the effective channel can be reduced to an addition of input noise followed by loss or  by a deterministic amplification, however we consider those two elements here to stay in a more general case. 

As shown in the appendix \ref{app_appen2}, an amplified coherent state $\ket{g_{\rm in} \gamma}\bra{g_{\rm in} \gamma}$ is therefore also transformed to a displaced thermal state
\begin{align}
\op{\sigma}_{\rm eff}(\gamma)=\op{D}(g_{\rm in}\sqrt{\tau  G}\gamma)\op{\rho}_{\rm th}(\lambda^g_{\rm ch})\op{D}^\dagger(g_{\rm in}\sqrt{\tau  G}\gamma)
\end{align}
of parameter $\lambda^g_{\rm ch}$ and mean amplitude $g_{\rm in} \sqrt{\tau G} \gamma$, leading to the output state
\begin{align}
\op{\rho}_{\rm out}^{\rm eff}=
\int \mathrm{d}^2\gamma \text{ } P_{\rm in}(\gamma)\op{\sigma}_{\rm eff}(\gamma)e^{(g_{\rm in}^2{-}1)\vert\gamma\vert^2}.
\label{NLA_canal_main}
\end{align}

The output states \eqref{etat_amplifie_apres} and \eqref{NLA_canal_main} will be proportional, with a state-independent factor, if the condition \eqref{condition_gin} is satisfied, and if $\op{\sigma}(\gamma)$ and $\op{\sigma}_{\rm eff}(\gamma)$ are equal, that is if
\begin{align}
g_{\rm in}\sqrt{\tau G}&=\tilde{g}\sqrt{T},  \\
\lambda_{\rm ch}^g&=g \lambda_{\rm ch}. 
\end{align}

The resolution of this set of equations gives the following effective parameters:
\begin{align}
g_{\rm in}&=  \sqrt{\frac{2+\left(g^2{-}1\right) \left(2{-}\epsilon\right) T}{2-\left(g^2{-}1\right)
   \epsilon T}} \label{eff_gin}\\
\tau  G & = \frac{g^2 T}{1+\left(g^2{-}1\right) T [\frac{1}{4} \left(g^2{-}1\right) \left(\epsilon{-}2\right) \epsilon T{-}\epsilon{+}1]}  := \eta \label{eff_eta} \\
\Delta &=\frac{2}{G}+\frac{ 2{-}\epsilon}{2} \left[\left(g^2{-}1\right) T \epsilon {-}2\right] \label{eff_delta}
\end{align}

\subsection{Properties of the effective channel}
Let us first comment some properties of the effective channel. First, there is only a condition on the product $\tau G{=}\eta$, and not on $\tau$ and $G$ separately. The input noise $\Delta$ also depends on $G$, since when $G$ increases, for given values of $\eta$ and of the output noise,  more noise is added by the deterministic amplification, and hence less input noise is needed.

\subsubsection{Added noise} 
There is a channel degeneracy: several combinations $(\Delta, G, \tau)$ can be equivalent to the same initial channel $\mathcal{L}$ followed by the real NLA. Indeed,  a state of variance $V$ is transformed to an output state of variance \footnote{We use the convention $N_0{=}1$ for the shot noise.}
\begin{subequations}
\begin{align}
V_{\rm out}&=\tau \Big[ G(V{+}\Delta)+ (G{-}1) \Big]+1{-}\tau, \\
&= \tau  G\left( V+\Delta{+}\frac{G{-}1}{G}{+}\frac{1{-}\tau }{\tau G}\right), \label{expression_variance}
\end{align}
\end{subequations}
and one can define a \textit{total} added noise at the input
\begin{align}
\chi_{\rm tot}=\Delta + \chi_{\rm ch}, \label{definition_chi_tot}
\end{align}
composed of the input noise $\Delta$, and of the noise due to the deterministic amplification and to the loss:
\begin{align}
\chi_{\rm ch}=\frac{G{-}1}{G}{+}\frac{1{-}\tau }{\tau G}=\frac{\tau (G{-}2){+}1}{\tau  G} \label{definition_chi_ch}
\end{align}

We stress that $g_{\rm in}$ defined by \eqref{eff_gin} does not depend on the choice of $\mathcal{C}_{\rm eff}$, as well as $\chi_{\rm tot}$:
\begin{align}
\chi_{\rm tot}=\frac{1}{g^2 T}+\frac{ \epsilon  \left[4{-}\left(g^4{-}1\right) T (\epsilon{-}2)\right]{-}4}{4
   g^2 }
\end{align}

\subsubsection{Three kinds of effective channels}
Using the effective channel degeneracy, one can find the simplest one, depending on the value of $\eta$:
\begin{itemize}
\item $\eta{\leq}1$: one can set $G{=}1$ and $\tau{=}\eta$. In that case, one recovers the effective parameters of \cite{blandino_improving_2012}, and the effective channel $\mathcal{C}_{\rm eff}$ is composed of a lossy channel of transmission $\eta$, with an input noise $\Delta_{G{=}1}{=}\epsilon^g$ and $\chi_{\rm ch}{=}\frac{1{-}\eta}{\eta}$.
\item $\eta{=}1$: one can set $G{=}\tau{=}1$, and the effective channel $\mathcal{C}_{\rm eff}$ is simply composed of an input noise addition $\Delta_{G{=}1}{=}\epsilon^g$.
\item $\eta{\geq}1$: one can set $\tau{=}1$ and $G{=}\eta$. In that case, the effective channel $\mathcal{C}_{\rm eff}$ is composed of a deterministic phase insensitive amplifier of gain  $G{=}\eta$, with an input noise $\Delta$ and $\chi_{\rm ch}{=}\frac{\eta{-}1}{\eta}$.
\end{itemize}

We stress that the effective parameters are obtained by a general method without involving any normalization, hence they are independent of the input state. The equivalence shown is this paper is also still valid if the input state has several modes, with one sent through the channel.

\subsection{Properties of the effective parameters}

Since the perfect NLA is theoretically described by an unbounded operator, it can lead to non-physical amplified states. For the same reason, it can lead to non-physical effective parameters when the gain of the real NLA is too large. Thus, the following constraints must be satisfied: the effective gain must be real and non divergent; each displaced thermal states  given by \eqref{displaced_thermal}  must not diverge; the global transmission $\eta$ must not diverge; and the input noise $\Delta$ must be positive.

Remarkably,  each of all those constraints leads to the same single condition on $g$, given by
\begin{align}
g<g_{\rm lim}= \sqrt{1+\frac{2}{T\epsilon}}. \label{condition_g}
\end{align}
As long as \eqref{condition_g} is satisfied, the effective parameters have a physical meaning. However, one has to be careful that this does not ensure that the amplified output state will be physical, as this depends on the input state. One can also define the maximum amount of noise for a given gain of the NLA from $g_{\rm lim}$:
\begin{align}
\epsilon_{\rm lim}=\frac{2}{(g^2{-}1)T}
\label{epsilon_lim}
\end{align}

When the channel is noiseless, i.e. when $\epsilon{=}0$, $g$ is no constrained by the effective channel, as pointed out by several prior studies (see e.g. \cite{ralph_nondeterministic_2008}).

As shown in \cite{blandino_improving_2012}, $\eta$ is smaller than 1 as long as $g$ is smaller than a value $g_{\rm max}$ which depends on $T$ and $\epsilon$. It is straightforward to see that $g_{\rm max}$ is always smaller than $g_{\rm lim}$, and therefore the physicality constraints are always fulfilled if the effective channel is restricted to a noisy and lossy channel.

Let us now highlight an important property of those effective parameters, coming from the fact that we consider the global transformation composed of the initial quantum channel and the NLA. Generally speaking, they increase with all parameters $g$, $T$, or $\epsilon$. In particular, as soon as $\epsilon{>}0$, $g_{\rm max}$ will not be infinite and there will be a value of $g$ such that $\eta{=}1$. On the contrary, for a fixed value of $g$, the value of $\eta$ increases with $\epsilon$. By adding thermal noise on purpose, it is thus possible to convert the initial channel to a lossless channel with $\eta{=1}$, for any gain $g$ of the NLA greater than 1. Naturally, the smaller the gain $g$, the greater the noise to add. This property will be analyzed in the next section.


\section{Application to quantum communications}

In this section, we present two applications of our results, for loss suppression and phase concentration. We note that we can also recover the results of \cite{blandino_improving_2012}, since when the input state is an EPR state of parameter $\lambda$, the effective NLA transforms it to another EPR state of parameter $g_{\rm in} \lambda$. If the gain of the NLA is smaller than $g_{\rm max}$, we can use the effective channel degeneracy and set $G{=}1$.

\subsection{Loss suppression}

M. Mi\v cuda \textit{et al.} have introduced the concept of noiseless attenuation, which allows to reduce the loss from a channel, when used with a NLA of appropriate gain \cite{micuda_noiseless_2012}. This attenuator is a NLA of gain $\nu{<}1$, which can be implemented by sending the state to be attenuated through a beam-splitter of amplitude transmission $\nu$, and conditioning on the vacuum for the reflected mode.

The principle of their protocol is the following: the initial state is first noiselessly attenuated with a factor $\nu$. It is then sent through the quantum channel, assumed noiseless in \cite{micuda_noiseless_2012}, which reduces its amplitude by $\sqrt{T}$. Finally, the state is noiselessly amplified with a NLA of gain $g{=}\frac{1}{\nu \sqrt{T}}$. In the limit $\nu{\to}0$, the protocol tends to the identity operation, and the input state does not undergo any loss. On the contrary, for a non zero $\nu$, the output state is also `contaminated' by noisy terms.

Apart from the fact that $\nu{=}0$ corresponds to an infinite value of $g$, and hence a zero probability of success, this suppression of loss does not take a simple form for $\nu{>}0$. The results of \cite{micuda_noiseless_2012} are also valid only for a noiseless channel. Using the equivalent system presented in this paper, the generalization of loss suppression is straightforward not only for a non zero $\nu$, but also for a noisy channel. As shown below, by using the appropriate gain for the noiseless attenuation, it is possible to exactly reduce loss, even if this gain does not tend to zero. 

\begin{figure}[t]
\includegraphics[width=0.97 \columnwidth]{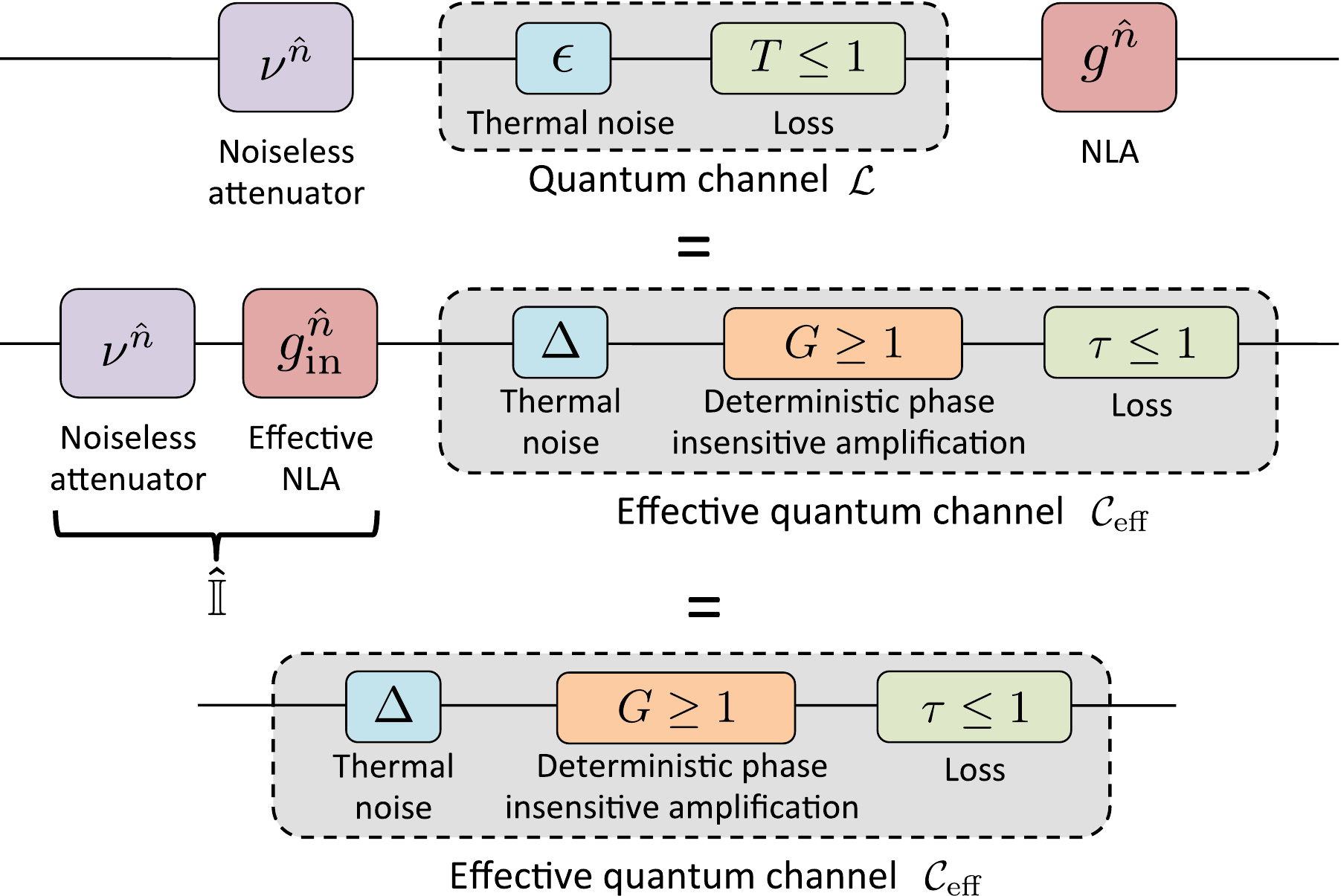}
\caption{(Color online) Loss suppression using a noiseless attenuator. For a noisy channel, a perfect loss suppression can be achieved with a finite gain $g$. }
\label{loss_suppression}
\end{figure}

Indeed, we have shown that a NLA after a quantum channel $\mathcal{L}$ is equivalent to an effective NLA of gain $g_{\rm in}$ before an effective channel $\mathcal{C}_{\rm eff}$. Therefore, an attenuator of gain $1/g_{\rm in}$ completely compensates the action of the effective NLA, since
\begin{align}
\left(1/g_{\rm in}\right)^{\hat{n}}g_{\rm in}^{\hat{n}}=\op{\mathbb{I}}.
\end{align}
There remains only the effective channel $\mathcal{C}_{\rm eff}$, as depicted in Fig. \ref{loss_suppression}.

For a noiseless channel ($\epsilon{=}0$), the effective gain is given by $g_{\rm in}{=}\sqrt{1{+}(g^2{-}1)T}$, and the effective parameter $\eta$, given by
\begin{align}
\eta=\frac{g^2 T}{1{+}(g^2{-}1)T},
\end{align}
always satisfies $T\leq\eta\leq1$ for $g\geq 1$. One can thus \textit{exactly} obtain a channel with smaller loss, using an attenuator of gain
\begin{align}
\nu=\frac{1}{g_{\rm in}}=\frac{1}{\sqrt{1{+}(g^2{-}1)T}}.
\label{gain_eff_sans_bruit}
\end{align}
For a gain $g>>1$, \eqref{gain_eff_sans_bruit} becomes
\begin{align}
\nu\simeq\frac{1}{g\sqrt{T}},
\end{align}
which corresponds to the gain used in \cite{micuda_noiseless_2012}. We see here that using a gain \eqref{gain_eff_sans_bruit} instead always leads to an exact channel with lower loss for any value of $g$.

When the initial channel is noisy, the effective channel can always be transformed to lossless channel with $\eta{=}1$. For a given value of $g$, this can be achieved by adding some noise between the attenuator and the channel, which allows to fully suppress loss, with a finite gain, but at the price of having more noise. 

\begin{figure}[t]
\subfloat[]{\includegraphics[width=0.94 \columnwidth]{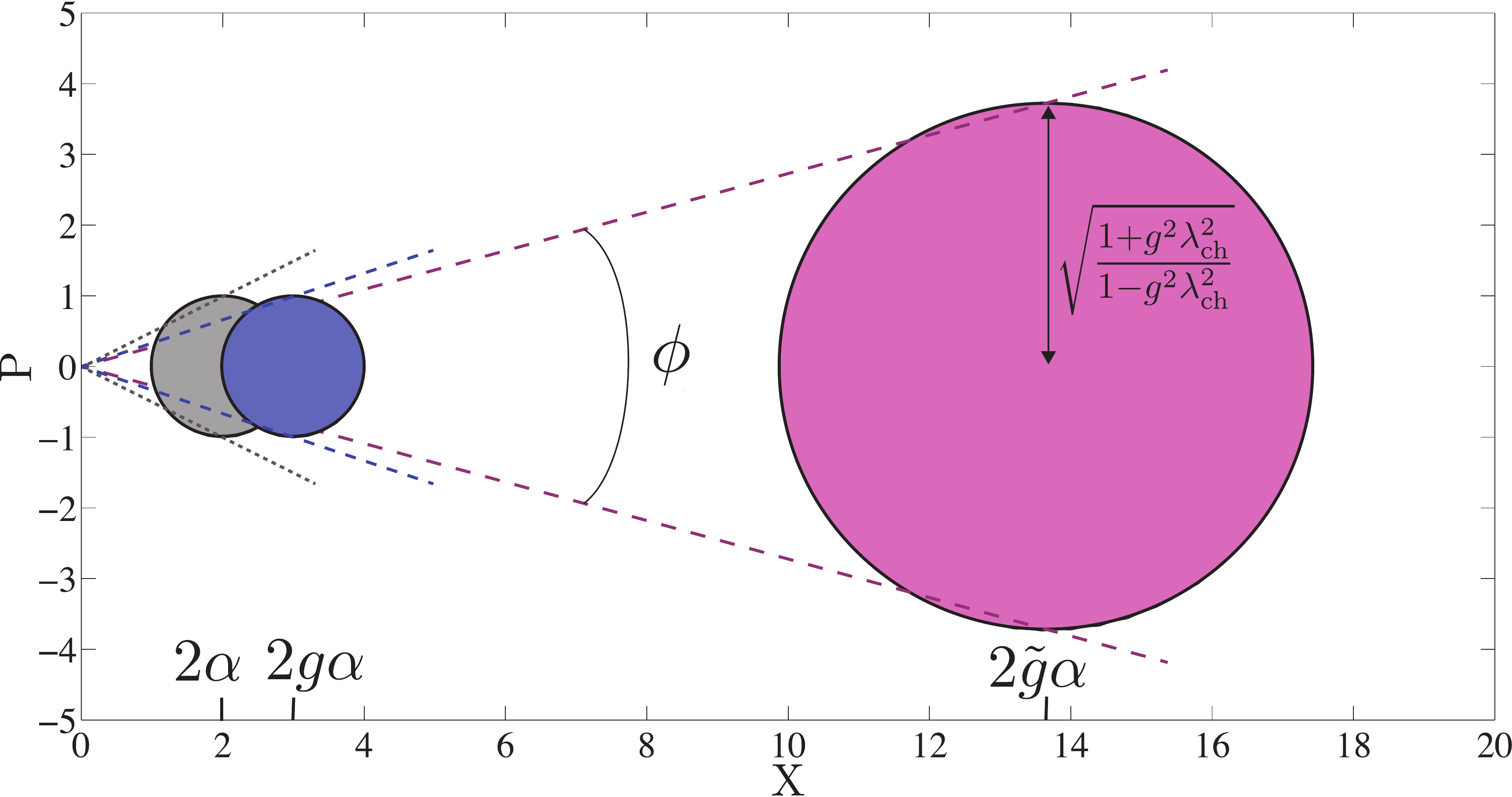}} \\
\subfloat[]{\includegraphics[width=5cm]{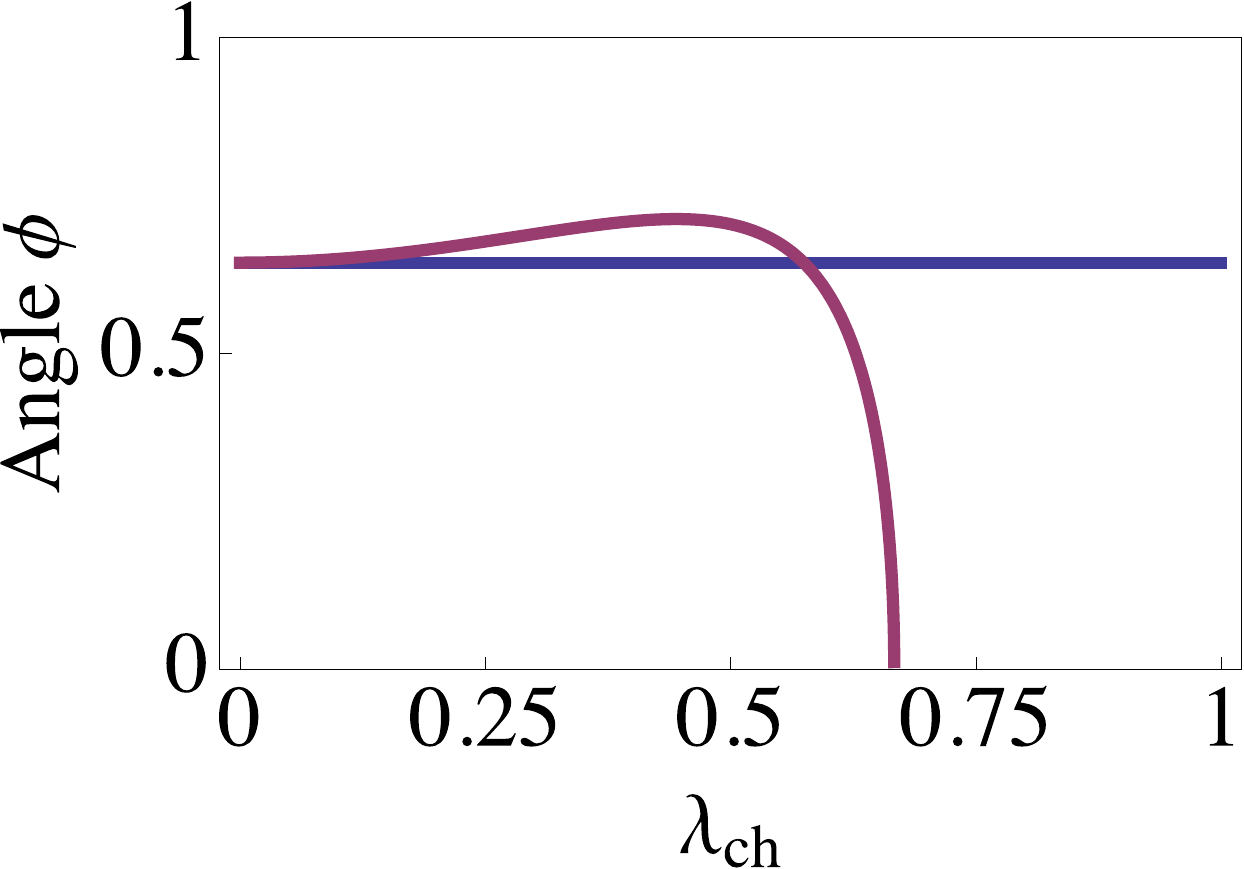}}
\caption{(Color online) Phase concentration by adding noise before a noiseless amplification. (a) Example of Wigner functions (with one standard deviation of radius) of the initial coherent state with $\alpha{=}1$, of the amplified state obtained with a NLA of gain $g{=}1.5$, and of the amplified state when thermal noise is added before the NLA. (b) Phase $\phi$ as defined by \eqref{phase_variance}. The input coherent state has an amplitude $\alpha{=}1$, and the NLA has a gain $g{=}1.5$.  Note that the factor 2 comes from the convention used for $N_0$. The blue curve is without thermal noise, and the pink curve is for a thermal noise $\epsilon$ defined by $\lambda_{\rm ch}^2{=}\frac{\epsilon}{2{+}\epsilon}$.}
\label{concentration_phase}
\end{figure}

\subsection{Phase concentration and SNR augmentation}

The experimental implementation of a NLA is very demanding on resources, especially on single-photon for most of the schemes. A protocol proposed by P. Marek and R. Filip  \cite{marek_coherent-state_2010} allows a particularly simple setup, as experimentally demonstrated by the group of U. Andersen \cite{usuga_noise-powered_2010}. The principle is the following: the coherent state is randomly displaced around its mean value, which corresponds to thermal noise addition. A photon is then subtracted from the noisy state. Although this scheme does not strictly produce an amplified coherent state, the photon subtraction will `select' high amplitudes $\ket{\beta}\bra{\beta}$ with a weight $\vert \beta \vert^2$, leading to a reduction of the phase variance, hence the appellation of phase concentration. 

Following the same idea, but replacing the photon subtraction by a NLA, high amplitudes coherent states will also be selected, but with an exponential factor $\exp[(g^2{-}1)\vert \beta \vert^2]$.  For a NLA of given gain $g$, it thus appears that adding noise before the noiseless amplification can also lead to phase concentration.  We use a simple criteria to define the phase uncertainty $\phi$ by
\begin{align}
\tan \frac{\phi}{2}=\frac{\text{Standard deviation}}{\text{Mean amplitude}}=\frac{\sqrt{\frac{1{+}g^2 \lambda^2_{\rm ch}}{1{-}g^2 \lambda^2_{\rm ch}}}}{2 \tilde{g} \alpha},
\label{phase_variance}
\end{align}
as shown in Fig. \ref{concentration_phase} (a). This quantity therefore corresponds to the inverse of the square root of the output SNR, and can be easily calculated using the equivalent system, since the noise addition corresponds to a channel with $T{=}1$. Note that there is no loss in the effective channel $\mathcal{C}_{\rm eff}$ in that case ($\tau{=}1$), since $\eta{>}1$ as soon as $T{=}1$ and $\epsilon{\ne}0$. In that picture, the effective NLA transforms the initial coherent state of mean amplitude $\alpha$ to a coherent state of mean amplitude $g_{\rm in} \alpha$.  The effective channel then degrades the SNR by adding the equivalent input noise $\chi_{\rm tot}{=}\Delta_{\tau{=}1}{+}\frac{\eta{-}1}{\eta}$ \eqref{definition_chi_tot}. The phase uncertainty \eqref{phase_variance} is therefore also given by
\begin{align}
\tan \frac{\phi}{2}=\frac{\sqrt{1{+}\chi_{\rm tot}}}{2 g_{\rm in} \alpha}.
\end{align}

Fig. \ref{concentration_phase} (b)  shows that $\phi$ can be theoretically reduced to an arbitrarily low value with the noise addition. The parameter $\lambda_{\rm ch}$ goes to the maximal corresponding value $\epsilon_{\rm lim}$. From this result, we can also conclude that the SNR can be arbitrarily increased, for a NLA of fixed gain, by adding thermal noise to the state.

Note also that the importance of thermal noise with noiseless amplification was also observed in \cite{blandino_improving_2012}, since the NLA does not improve the key rate when the channel has loss only.


\section{Discussion and conclusion}

We have discussed a general equivalence when a NLA is used after a noisy and lossy quantum channel, and shown that the transformation is equal to an effective NLA followed by an effective channel, up to a state-independent proportionality factor. This equivalence is valid regardless of the nature or the input state, which may be non-Gaussian, or part of a multimode state. 

Using this picture, we have analyzed several applications: when a suitable noiseless attenuation is used before the quantum channel, it is possible to obtain an exact effective quantum channel with smaller loss, but with larger noise if the initial noise is non zero. Increasing the gain of the NLA, or deliberately adding noise before the quantum channel (and after the noiseless attenuation) can lead to a perfectly noisy lossless quantum channel. We have also shown that this noise addition can be used to reduce the phase uncertainty of input coherent states. 

As shown with those two applications, our results not only allow for a simpler calculation of the amplified states, but they also provide a detailed physical explanation, which is likely to be useful for future applications.

\section{Acknowledgements}
R.B. would like to thank T.C. Ralph, N. Walk and A.P. Lund for useful discussions. This research was partially funded by the Australian Research Council Centre of Excellence for Quantum Computation and Communication Technology (Project No. CE11000102). We acknowledge support from the ERA-Net project HIPERCOM. M.B. is partially supported by a Rita Levi-Montalcini contract of MIUR and by the EPSRC Programme Grant EP/K034480/1.


%

\appendix



\section{Amplification after the quantum channel}
\label{appen1}
In this appendix, we detail the derivation of the effective system. Let us begin by studying the action of a NLA placed after a linear symmetric lossy and noisy Gaussian quantum channel $\mathcal{L}$, as pictured on Fig. \ref{ampli_deterministe_phaseIndependant} (a). This channel has a transmission $T$, and an input noise $\epsilon$. For the sake of simplicity, one can consider that such a channel is composed of the addition of thermal noise $\epsilon$ at its input, followed by a lossy noiseless channel of transmission $T$. An input state having a quadrature variance $V$ is thus transformed to a state of variance  $T(V{+}\epsilon){+}1{-}T$.

We associate an operation $\mathbf{\mathfrak{L}}$ to this quantum channel. Since the amplification of a coherent state is simply given by  \eqref{amplification_coherent_state}, the $P$ function is a very useful tool to compute the amplification of an arbitrary state. 

\subsection{Action of $\mathcal{L}$} Let us consider an arbitrary quantum state given by \eqref{decomposition_rho_in_NLA}. Using the linearity of $\mathfrak{L}$, the output state of the channel, before the NLA, is given by
 \begin{align}
 \op{\rho}_{\rm out}{=}\mathfrak{L}[\op{\rho}_{\rm in}]{=}\int \mathrm{d}^2\gamma \text{ } P_{\rm in}(\gamma)\mathfrak{L}[\ket{\gamma}\bra{\gamma}]. \label{etat_sortie_canal}
 \end{align}
The NLA then produces an (unnormalized) amplified state $\op{\rho}_{\rm out}^{\rm NLA}$:
\begin{subequations}
\begin{align}
\op{\rho}_{\rm out}^{\rm NLA}&=\op{T}\op{\rho}_{\rm out}\op{T}\\
&=\int \mathrm{d}^2\gamma \text{ } P_{\rm in}(\gamma)\op{T}\mathfrak{L}[\ket{\gamma}\bra{\gamma}]\op{T}  \label{etat_quelconque_amplifie_nonDetail}
\end{align}
\end{subequations} 

Therefore, due to the channel linearity, it is sufficient to know the evolution of a coherent state $\ket{\gamma}\bra{\gamma}$ in order to obtain the evolution of an arbitrary state. The transformation of a coherent state by the lossy and noisy channel is trivial: first, the mean amplitude $\gamma$ is transformed to $\sqrt{T}\gamma$. Then, the variance of the quadratures is transformed to $T(1{+}\epsilon){+}1{-}T{=}1{+}T\epsilon$. Since the channel is assumed to be symmetric and Gaussian, the state $\mathfrak{L}\left[\ket{\gamma}\bra{\gamma}\right]$ is therefore a thermal state $\op{\rho}_{\rm th}(\lambda_{\rm ch})$ displaced by $\sqrt{T}\gamma$:
\begin{align}
\mathfrak{L}\left[\ket{\gamma}\bra{\gamma}\right]=\op{D}(\sqrt{T}\gamma)\op{\rho}_{\rm th}(\lambda_{\rm ch})\op{D}^\dagger(\sqrt{T}\gamma)
\end{align}
The parameter $\lambda_{\rm ch}$ is such that the variance of $\op{\rho}_{\rm th}(\lambda_{\rm ch})$ equals $1{+}T\epsilon$, which gives
\begin{align}
\lambda_{\rm ch}^2=\frac{T\epsilon}{2{+}T\epsilon}.
\label{lambda_ch}
\end{align} 

\subsection{Amplification of a displaced thermal state}

In order to compute the action of the NLA, one can also express the displaced thermal state $\mathfrak{L}\left[\ket{\gamma}\bra{\gamma}\right]$ using the $P$ function:
\begin{align}
\mathfrak{L}\left[\ket{\gamma}\bra{\gamma}\right]=\int \mathrm{d}^2\alpha \text{ } P_{\gamma}(\alpha)\ket{\alpha}\bra{\alpha}
\end{align}
As shown in \cite{blandino_improving_2012}, $P_{\gamma}(\alpha)$ can be expressed as
\begin{align}
P_\gamma(\alpha)=P_{\gamma_x}(\alpha_x)P_{\gamma_y}(\alpha_y), 
\end{align}
where 
\begin{subequations}
\begin{align} 	
P_{\gamma_x}(\alpha_x)=\frac{1}{\sqrt{\pi}}\sqrt{\frac{1{-}\lambda_{\rm ch}^2}{\lambda_{\rm ch}^2}}e^{-\frac{1{-}\lambda_{\rm ch}^2}{\lambda_{\rm ch}^2}(\alpha_x{-}\sqrt{T}\gamma_x)^2}, \\
P_{\gamma_y}(\alpha_y)=\frac{1}{\sqrt{\pi}}\sqrt{\frac{1{-}\lambda_{\rm ch}^2}{\lambda_{\rm ch}^2}}e^{{-}\frac{1{-}\lambda_{\rm ch}^2}{\lambda_{\rm ch}^2}(\alpha_y{-}\sqrt{T}\gamma_y)^2}.	
	\end{align}
	\end{subequations}

Using again the linearity of the NLA, the amplification of a displaced thermal state is given by:
\begin{subequations}
\begin{align}
\op{T}\mathfrak{L}[\ket{\gamma}\bra{\gamma}]\op{T}&=\int \mathrm{d}^2\alpha \text{ } P_{\gamma}(\alpha)\op{T}\ket{\alpha}\bra{\alpha}\op{T}\\
&=\int \mathrm{d}^2 \alpha \text{ } P_\gamma(\alpha)e^{(g^2{-}1)\vert \alpha \vert ^2}\ket{g\alpha}\bra{g\alpha}
\end{align}
\end{subequations}
Then, the change of variable $u{=}g\alpha{=}u_x{+}iu_y$ gives $\mathrm{d}^2\alpha{=}\mathrm{d}^2u/g^2$, and
\begin{align}
\op{T}\mathfrak{L}[\ket{\gamma}\bra{\gamma}]\op{T}=\int \frac{1}{g^2}\mathrm{d}^2u \text{ } P_\gamma(u/g)e^{\frac{g^2{-}1}{g^2}\vert u \vert^2}\ket{u}\bra{u}
\end{align}
As before, one can separate the variables $u_x$ and $u_y$. We now focus on $u_x$, the results being similar for $u_y$. We first highlight that
\begin{subequations}
\begin{align}
&\frac{1}{g}P_{\gamma_x}(u_x/g)e^{\frac{g^2{-}1}{g^2}u_x^2} \\
&=\frac{1}{g}\frac{1}{\sqrt{\pi}}\sqrt{\frac{1{-}\lambda_{\rm ch}^2}{\lambda_{\rm ch}^2}}e^{{-}\frac{1{-}\lambda_{\rm ch}^2}{\lambda_{\rm ch}^2}(\frac{u_x}{g}{-}\sqrt{T}\gamma_x)^2{+}\frac{g^2{-}1}{g^2}u_x^2}\\
&=\sqrt{\frac{1{-}\lambda^2_{\rm ch}}{1{-}g^2\lambda^2_{\rm ch}}}\frac{1}{\sqrt{\pi}}\sqrt{\frac{1{-}g^2\lambda_{\rm ch}^2}{g^2\lambda_{\rm ch}^2}}e^{{-}\frac{1{-}\lambda_{\rm ch}^2}{\lambda_{\rm ch}^2}(\frac{u_x}{g}{-}\sqrt{T}\gamma_x)^2{+}\frac{g^2{-}1}{g^2}u_x^2}.
\end{align}
\end{subequations}
The argument of the exponential can be easily put in the form
\begin{align}
-\frac{1{-}\lambda_{\rm ch}^2}{\lambda_{\rm ch}^2}\Big(\frac{u_x}{g}{-}&\sqrt{T}\gamma_x\Big)^2+\frac{g^2{-}1}{g^2}u_x^2 \\
&=\underbrace{{-}\frac{1{-}g^2\lambda_{\rm ch}^2}{g^2\lambda_{\rm ch}^2}}_{\substack{\text{Thermal state} \\ \text{of parameter $g\lambda_{\rm ch}$}}}\Big(u_x{-}\sqrt{T}\gamma_x \underbrace{g\frac{1{-}\lambda_{\rm ch}^2}{1{-}g^2\lambda_{\rm ch}^2}}_{\text{Gain}}\Big)^2 \\
&+\underbrace{T\gamma_x^2\frac{(g^2{-}1)(1{-}\lambda_{\rm ch}^2)}{1{-}g^2\lambda_{\rm ch}^2}}_{\substack{\text{Normalization term}\\\text{independant of $u_x$}}}.
\end{align}

Apart from the normalization term, one easily recognizes the signature of a thermal state of parameter $g\lambda_{\rm ch}$ and of variance
\begin{align}
\frac{1+g^2\lambda^2_{\rm ch}}{1-g^2 \lambda^2_{\rm ch}}=\frac{2+T \epsilon(1{+}g^2)}{2+T \epsilon(1{-}g^2)},
\end{align}
displaced by $g\frac{1{-}\lambda_{\rm ch}^2}{1{-}g^2\lambda_{\rm ch}^2}\sqrt{T}\gamma_x$. The NLA thus amplifies the mean amplitude of the state with a gain
\begin{align}
\tilde{g}=g\frac{1{-}\lambda_{\rm ch}^2}{1{-}g^2\lambda_{\rm ch}^2} \label{tilde_g_appendix}
\end{align}
greater than $g$, since $g\lambda_{\rm ch}$ must remain smaller than 1 for the amplified state to be physical.

\begin{widetext}
In conclusion, the (unnormalized) amplification of a displaced thermal state is given by
\begin{align}
\op{T}\mathfrak{L}[\ket{\gamma}\bra{\gamma}]\op{T}= \op{D}\Big(\tilde{g}\sqrt{T}\gamma\Big)\op{\rho}_{\rm th}(g\lambda_{\rm ch})\op{D}^\dagger\Big(\tilde{g}\sqrt{T}\gamma\Big)\times \left(\frac{1{-}\lambda^2_{\rm ch}}{1{-}g^2\lambda^2_{\rm ch}}\right) e^{T \vert \gamma \vert^2 \frac{(g^2{-}1)(1{-}\lambda_{\rm ch}^2)}{1{-}g^2\lambda_{\rm ch}^2}}.
\label{etat_thermique_deplace_amplifie}
\end{align}

\end{widetext}

Finally, by inserting \eqref{etat_thermique_deplace_amplifie} in \eqref{etat_quelconque_amplifie_nonDetail}, the amplified state produced by the system depicted in Fig. \ref{ampli_deterministe_phaseIndependant} (a) is given by
\begin{align}
&\op{\rho}_{\rm out}^{\rm NLA} = \left(\frac{1{-}\lambda^2_{\rm ch}}{1{-}g^2\lambda^2_{\rm ch}}\right)\int \mathrm{d}^2\gamma \text{ } P_{\rm in}(\gamma)  \op{\sigma}(\gamma) e^{\vert \gamma \vert^2 T \frac{(g^2{-}1)(1{-}\lambda_{\rm ch}^2)}{1{-}g^2\lambda_{\rm ch}^2}},
\label{rho_out_nla}
\end{align}
where
\begin{align}
\op{\sigma}(\gamma)=\op{D}\big(\tilde{g}\sqrt{T}\gamma\big)\op{\rho}_{\rm th}(g\lambda_{\rm ch})\op{D}^\dagger\big(\tilde{g}\sqrt{T}\gamma\big).
\end{align}

\section{Effective system}
\label{app_appen2}
\subsection{Amplification by the effective NLA}
The effective channel $\mathcal{C}_{\rm eff}$ following the effective NLA is described by an operation $\mathfrak{L}_{g}$. As explained in the main text,  we look for parameters $g_{\rm in}$, $\Delta$, $G$, and $\tau $ such that
\begin{align}
\op{T}\mathfrak{L}[\op{\rho}_{\rm in}]\op{T}=\mu\text{ } \mathfrak{L}_g[\op{T}_{\rm in}\op{\rho}_{\rm in} \op{T}_{\rm in}],
\label{nla_effectif_egalite_reel}
\end{align}
where $\op{T}_{\rm in}{=}g_{\rm in}^{\hat{n}}$ is the operator associated to the effective NLA, and $\mu$ is a constant factor, independent of $\op{\rho}_{\rm in}$. 

Let us start by writing the noiseless amplification of $\op{\rho}_{\rm in}$, using the $P$ function:
\begin{subequations}
\begin{align}
\op{T}_{\rm in}\op{\rho}_{\rm in}\op{T}_{\rm in}&=\int \mathrm{d}^2\gamma \text{ } P_{\rm in}(\gamma)\op{T}_{\rm in}\ket{\gamma}\bra{\gamma}\op{T}_{\rm in} \\
&=\int \mathrm{d}^2\gamma \text{ } P_{\rm in}(\gamma)\ket{g_{\rm in}\gamma}\bra{g_{\rm in}\gamma}e^{(g_{\rm in}^2{-}1)\vert\gamma\vert^2}
\label{NLAeffectif_amplification}
\end{align} 
\end{subequations}

\subsection{Output state after the effective channel}

Since  $\mathcal{C}_{\rm eff}$ is a symmetric and Gaussian channel, a coherent state $\ket{g_{\rm in}\gamma}$ is simply transformed to a state of mean amplitude $ g_{\rm in} \sqrt{\tau  G}\gamma$, with a variance 
\begin{subequations}
\begin{align}
V_{\rm out}&=\tau\Big[G(1{+}\Delta){+}G{-}1\Big]{+}1{-}\tau\\
&=1+ \tau  G \Delta {+}2\tau  (G{-}1)
\end{align} 
\end{subequations}
for both quadratures. It can thus be written as a displaced thermal state $\op{\sigma}_{\rm eff}(\gamma)=\op{D}(g_{\rm in}\sqrt{\tau  G}\gamma)\op{\rho}_{\rm th}(\lambda^g_{\rm ch})\op{D}^\dagger(g_{\rm in}\sqrt{\tau  G}\gamma)$, where $\lambda^g_{\rm ch}$ is such that  $V_{\rm out}{=}\frac{1{+}(\lambda^g_{\rm ch})^2}{1{-}(\lambda^g_{\rm ch})^2}$, which gives
\begin{align}
 \lambda^{g}_{\rm ch}{=}\sqrt{\frac{\tau   (\Delta  G{+}2 G{-}2)}{ \Delta  \tau   G{+}2 \tau   G{+}2(1{-}\tau )}}.
\end{align}
After the effective quantum channel $\mathcal{C}_{\rm eff}$, \eqref{NLAeffectif_amplification} finally becomes
\begin{align}
\mathfrak{L}_g[\op{T}_{\rm in}\op{\rho}_{\rm in} \op{T}_{\rm in}]=
\int \mathrm{d}^2\gamma \text{ } P_{\rm in}(\gamma)\op{\sigma}_{\rm eff}(\gamma)e^{(g_{\rm in}^2{-}1)\vert\gamma\vert^2}.
\label{NLA_canal}
\end{align}

\subsection{Conditions for the effective parameters}
Comparing the states  \eqref{rho_out_nla}   and \eqref{NLA_canal}, one can identify a set of equations for the effective parameters. The first condition is given by comparing the exponential factors:
\begin{align}
g^2_{\rm in}{-}1&=T \frac{(g^2{-}1)(1{-}\lambda_{\rm ch}^2)}{1{-}g^2\lambda_{\rm ch}^2}
\label{condition_parametresEff2}
\end{align}
The second and third conditions are given by comparing the displaced thermal states $\op{\sigma}(\gamma)$ and $\op{\sigma}_{\rm eff}(\gamma)$, and by imposing the same mean amplitude and variance: 
\begin{align}
g_{\rm in}\sqrt{\tau G}&=\tilde{g}\sqrt{T} \label{condition_parametresEff1} \\
\lambda_{\rm ch}^g&=g \lambda_{\rm ch} \label{condition_lambdaTh} 
\end{align}

One can easily solve this system of equations, obtaining the effective parameters:
\begin{align}
		g_{\rm in}&=  \sqrt{\frac{2+\left(g^2{-}1\right) \left(2{-}\epsilon\right) T}{2-\left(g^2{-}1\right)
   \epsilon T}} \\
\tau  G & = \frac{g^2 T}{1+\left(g^2{-}1\right) T [\frac{1}{4} \left(g^2{-}1\right) \left(\epsilon{-}2\right) \epsilon T{-}\epsilon{+}1]}  := \eta\\
\Delta &=\frac{2}{G}+\frac{ 2{-}\epsilon}{2} \left[\left(g^2{-}1\right) T \epsilon {-}2\right]
\end{align}
The constant factor $\mu$ is given by
\begin{align}
\mu{=}\frac{1{-} \lambda^2_{\rm ch}}{1{-}g^2 \lambda_{\rm ch}^2}, \label{definition_mu}
\end{align}
which is independent of the input state $\op{\rho}_{\rm in}$.

\end{document}